\documentstyle[11pt]{article}

\def\roughly#1{\mathrel{\raise.3ex\hbox{$#1$\kern-.75em%
\lower1ex\hbox{$\sim$}}}}

\def\lsim{\roughly<}
\def\gsim{\roughly>}
\def\be{\begin{eqnarray}}
\def\ee{\end{eqnarray}}
\def\susc{susceptibility}
\def\Tr{{\rm Tr}}

\def\ben{\begin{enumerate}}
\def\een{\end{enumerate}}
\def\beitem{\begin{itemize}}
\def\eitem{\end{itemize}}

\newcommand{\beq}{\begin{eqnarray}}
\newcommand{\eeq}{\end{eqnarray}}
\def\la{\langle}
\def\ra{\rangle}
\def\bi{\begin{itemize}}
\def\ei{\end{itemize}}
\def\ie{{\it i.e}}
\def\eg{{\it e.g.}}
\def\etal{{\it et al}}
\def\del{\partial}
\def\L{{\cal L}}

\long\def\beginomit#1\endomit{}
\def\np{{Nucl. Phys.}}
\def\prl{Phys. Rev. Lett.}
\def\pr {Phys. Rev.}
\def\PR {Phys. Repts.}

\def\pl{Phys. Lett.}
\def\L{{\cal L}}

\begin{document}


\begin{titlepage}\begin{center}

\hfill{SUNY NTG-94-44}

\hfill{hep-ph/9408223}

\vskip 0.4in
{\Large\bf CHIRAL SYMMETRY RESTORATION}
\vskip 0.1cm
{\Large\bf AND THE GEORGI VECTOR LIMIT}
\vskip 1.2in
{\large  G. E. Brown$^a$ and Mannque Rho$^b$}\\
\vskip 0.1in
{\large a) \it Department of Physics, State University of New York,} \\
{\large \it Stony Brook, N.Y. 11794, USA. }\\
{\large b) \it Service de Physique Th\'{e}orique, CEA  Saclay}\\
{\large\it 91191 Gif-sur-Yvette Cedex, France}\\
\vskip .6in
\centerline{July 1994}
\vskip .6in

{\bf ABSTRACT}\\ \vskip 0.1in
\begin{quotation}

\noindent  The chiral phase transition at high temperature and/or density
is interpreted in terms of Georgi's vector limit. We discuss three cases
as possible support for this scenario: Quark-number susceptibility, cool
kaons in heavy-ion process and an instanton-molecule picture for chiral
restoration. Both the notion of ``mended symmetry" and the Georgi vector limit
are suggested to be relevant in nuclear physics of dense matter.
\end{quotation}
\end{center}\end{titlepage}


\indent

One of the most intriguing -- and largely unsettled -- problems in strong
interaction physics is how chiral symmetry is
restored in matter as  it becomes hot and/or dense. In this letter, we propose
that the symmetry restoration involves the ``vector limit" put forward
by Georgi some years ago\cite{georgi}. Georgi envisaged the vector limit
to be appropriate in the large
$N_c$ limit where $N_c$ is the number of colors and
supposed that Nature with $N_c=3$ is close to the limit, so that the
departure from the limit could be treated as a perturbation. Here we
are proposing that the vector limit is appropriate in QCD at the chiral
phase transition in hot and/or dense hadronic matter.
As support to our proposal, we
shall discuss three cases: 1) lattice gauge calculations on quark-number
susceptibility at high temperature; 2) ``cool" kaons observed in
heavy-ion collisions; 3) an instanton-molecule model for chiral phase
transition.

\subsubsection*{The Georgi vector limit}
\indent

We begin by sketching the essential points of the Georgi vector symmetry
and vector limit\cite{georgi}. Consider two chiral flavors u(p) and d(own),
with chiral symmetry $SU(2)_L\times SU(2)_R$. The standard way of looking at
this symmetry is that it is realized either in Nambu-Goldstone (or Goldstone
in short) mode, with $SU(2)_L\times SU(2)_R$ broken down spontaneously to
$SU(2)_{L+R}$ or in Wigner-Weyl (or Wigner in short) mode with parity
doubling. Georgi observes, however, that
there is yet another way of realizing the symmetry which requires
both Goldstone mode and Wigner mode to {\it co-exist}. Now the signature
for {\it any} manifestation of the chiral symmetry is the pion decay constant
$f_\pi$
\be
\la 0 |A^i_\mu|\pi^j (q)\ra=i f_\pi q_\mu \delta^{ij}\label{goldstone}
\ee
where $A_\mu^i$ is the isovector axial current. The Goldstone mode is
characterized by the presence of the triplet of Goldstone bosons,
$\pi^i$ with $i=1, 2, 3$ with a non-zero pion decay constant.
The Wigner mode is realized when the pion decay constant vanishes, associated
with
the absence of zero-mass bosons. In the latter case, the symmetry is realized
in a parity-doubled mode. The Georgi vector symmetry we are interested in
corresponds to
the mode (\ref{goldstone}) co-existing with a triplet of scalars $S^i$
with $f_S=f_\pi$ where
\be
\la 0|V_\mu^i| S^j (q)\ra= i f_S q_\mu \delta^{ij}
\ee
where $V_\mu^i$ is the isovector-vector current. In this case, the
$SU(2)\times SU(2)$ symmetry is unbroken. At low $T$ and/or low density,
low-lying isovector-scalars are {\it not} visible and hence either the vector
symmetry is broken in Nature with $f_S\neq f_\pi$ or they are ``hidden"
in the sense that they are eaten up by vector particles (\`a la Higgs).
In what follows, we would like to suggest that as temperature and/or density
rises to the critical value corresponding to the chiral phase transition,
the symmetry characterized by
\be
f_S=f_\pi\label{equal}
\ee
is restored with the isovector scalars making up the longitudinal components
of the massive $\rho$ mesons, which eventually get ``liberated" at some high
temperature (or density) from the
vectors and become degenerate with the zero-mass
pions at $T\gsim T_{\chi SR}$ where
$T_{\chi SR}\sim 140$ MeV is the chiral transition temperature.
The symmetry (\ref{equal}) with the scalars ``hidden" in the massive vector
mesons resembles Weinberg's mended symmetry presumed to be realized near the
chiral symmetry restoration\cite{weinbergMS}, so we shall
refer to this as ``mended symmetry." We shall reserve
``Georgi vector limit" as the symmetry limit in which (\ref{equal}) holds
together with $m_\pi=m_S=0$.

The relevant Lagrangian to use for illustrating
what we mean is the hidden gauge
symmetric Lagrangian of Bando {\etal}\ \cite{bando} which is valid below
the chiral transition\footnote{We will ignore the small quark masses and
work in the chiral limit, unless otherwise noted.},
\be
\L=\frac 12 f^2\left\{\Tr(D^\mu\xi_L D_\mu \xi_L^\dagger) +
(L\rightarrow R)\right\} +\kappa \cdot \frac 14 f^2 \Tr (\del^\mu U\del_\mu
U^\dagger) +\cdots\label{bandoL}
\ee
where
\be
U&=&\xi_L \xi_R^\dagger,\nonumber\\
D^\mu \xi_{L,R}&=&\del^\mu\xi_{L,R}-ig\xi_{L,R}\rho^\mu, \nonumber\\
\rho_\mu&\equiv& \frac 12 \tau^a \rho^a_\mu
\ee
and $g$ stands for the hidden gauge coupling. The ellipsis stands for other
matter fields and higher-derivative terms needed to make the theory more
realistic.
The $\xi$ field can be parametrized as
\be
\xi_{L,R}\equiv e^{iS(x)/f_S} e^{\pm i\pi (x)/f_\pi}
\ee
with $S(x)=\frac 12 \tau^a S^a (x)$ and $\pi (x)=\frac 12 \tau^a \pi^a (x)$.
Under the global chiral $SU(2)_L\times SU(2)_R$ transformation,
\be
\xi_L\rightarrow L\xi_L G^\dagger, \ \ \ \xi_R\rightarrow R\xi_R G^\dagger
\ee
with $L(R)\in SU(2)_{L(R)}$ and $G\in SU(2)_{local}$ is the hidden
local transformation. The Lagrangian (\ref{bandoL}) is invariant under
$G$. Thus the symmetry of the Lagrangian (\ref{bandoL}) is
$(SU(2)_L\times SU(2)_R)_{global} \times G_{local}$.
Setting $S(x)=0$ corresponds to taking the unitary gauge in which case
we are left with physical fields only ({\ie}, no ghost fields).

At tree level, we get that
\be
f_S=f, \ \ f_\pi=\sqrt{1+\kappa} f
\ee
and the $\rho\pi\pi$ coupling
\be
g_{\rho\pi\pi}=\frac{1}{2(1+\kappa)} g.
\ee
Going to the unitary gauge, one gets the KSRF mass relation
\be
m_\rho=fg=\frac{1}{\sqrt{1+\kappa}} f_\pi g =
2\sqrt{1+\kappa} f_\pi g_{\rho\pi\pi}.\label{KSRF}
\ee
Now we know from experiments that at zero $T$ (or low density), the
$\kappa$ takes the value $-\frac 12$ for which the KSRF relation is
accurately satisfied. The symmetry (\ref{equal}) therefore is broken.
The symmetry is recovered
for $\kappa=0$ in which case the second term of (\ref{bandoL})
that mixes L and R vanishes, thus restoring $SU(2)\times SU(2)$. In this
limit, the hidden gauge symmetry swells to $G_L\times G_R$, and $\xi_{L,R}$
transform
\be
\xi_L\rightarrow L\xi_L G_L^\dagger, \ \ \xi_R\rightarrow R\xi_R G_R^\dagger.
\ee
If the gauge coupling is not zero, then the $\rho$ mesons are still
massive and we have the mended symmetry (\ref{equal}). However if the
gauge coupling vanishes, then the vector mesons become massless and
their longitudinal components get liberated, giving the scalar massless
multiplets $S(x)$. In this limit, the symmetry is the global $[SU(2)]^4$.
Local symmetry is no longer present.

We shall now argue that in hot and/dense matter approaching the chiral
restoration, the constant $\kappa\rightarrow 0$ and the gauge coupling
$g\rightarrow 0$.  For this purpose, we shall extrapolate a bit
the results obtained by Harada and Yamawaki\cite{harada}. These authors
studied the hidden gauge Lagrangian (\ref{bandoL}) to one loop order
and obtained the $\beta$ functions for the hidden gauge coupling $g$
and the constant $\kappa$ (in dimensional regularization)
\be
\beta_g (g)&=& \mu \frac{dg}{d\mu}=-\frac{87-a^2}{12} \frac{g^2}{(4\pi)^2},
\\
\beta_\kappa (a)&=& \mu \frac{da}{d\mu}= 3a (a^2-1) \frac{g^2}{(4\pi)^2}
\ee
with $a=\frac{1}{1+\kappa}$. One notes that first of all, there
is a nontrivial ultraviolet fixed point at $a=1$ or $\kappa=0$
and that the coupling constant $g$ scales to zero as $\sim 1/\ln \mu$ in the
ultraviolet limit. This perturbative result may not be realistic
enough to be taken seriously -- and certainly cannot be pushed too high in
energy-momentum scale but for the reason given below, we think it
plausible that the Harada-Yamawaki results hold at least qualitatively as
high $T$ (or density) is reached.
In fact we expect that the gauge coupling should fall off to zero much faster
than logarithmically in order to explain what follows below.

\subsubsection*{Quark-number \susc}
\indent

As a first case for the scenario sketched above, we consider the
lattice gauge calculations by Gottlieb {\etal}
\cite{gottliebchi} of the quark-number susceptibility defined by
\be
\chi_{\pm}=\left(\del/\del \mu_{u} \pm \del/\del \mu_d\right) (\rho_u\pm
\rho_d)
\ee
where the $+$ and $-$ signs define the singlet (isospin zero) and triplet
(isospin one) susceptibilities, $\mu_u$ and $\mu_d$ are the chemical
potentials of the up and down quarks and
\be
\rho_i=\Tr N_i exp\left[-\beta (H-\sum_{j=u,d} \mu_j N_j)\right]/V
\equiv \la\la N_i\ra\ra/V
\ee
with $N_i$ the quark number operator for flavor $i=u,d$.
One can see that the $\chi_+$ is in the $\omega$-meson channel and
the $\chi_-$ in the $\rho$-meson channel.
For $SU(2)$ symmetry, we expect $\chi_+=\chi_-$ and this is what one observes
in the lattice results. One can classify the lattice results by roughly
three temperature regimes.  Focusing on the non-singlet susceptibility,
we see that in the very low temperature regime, the $\chi_-$ is dominated
by the $\rho$ meson and is small. As the temperature moves toward the onset of
the phase transition, the $\chi_-$ increases rapidly to near that of
non-interacting quarks. This regime may be described in terms of constituent
quarks. In RPA approximation of the constituent quark model as used
by Kunihiro\cite{kunihiro}, the susceptibility below the critical
temperature is
\be
\chi=\frac{\chi_0}{1+G_v \chi_0}
\ee
where $G_v$ is the coupling of the constituent quark (denoted $Q$) to the
vector meson $\rho$ and $\chi_0$ is the susceptibility for non-interacting
quarks which at $T\approx T_{\chi SR}$ where the dynamical mass $m_Q$ has
dropped to zero has the value
\be
\chi_0\approx N_f T^2
\ee
with $N_f$ the number of flavors. In terms of the gauge coupling of
(\ref{bandoL}), we have
\be
G_v\approx \frac{g^2}{4m_\rho^2}.
\ee
As noted by Kunihiro in the NJL model, the rapid increase of the
\susc\  can be understood by a steep drop in the vector coupling across
the $T_{\chi SR}$. Let us see what we obtain with the hidden gauge
symmetry Lagrangian (\ref{bandoL}). If we assume that the KSRF
relation (\ref{KSRF}) holds at $T$ near $T_{\chi SR}\approx 140$ MeV
(the recent work by Harada {\etal}\ \cite{haradaPRL} supports this
assumption) and that $\chi_0\approx 2T_{\chi SR}^2$ for $N_f=2$, then we find
\be
\chi (T_{\chi SR})/\chi_0 (T_{\chi SR})\approx \frac{1}{1+\frac 12
(\frac{T_{\chi SR}}{f_\pi})^2} \approx 0.47\label{HGSchi}
\ee
with $\kappa=0$.
Here we are assuming that $f_\pi$ remains at its zero temperature value,
93 MeV, up to near $T_{\chi SR}$. The ratio (\ref{HGSchi})
is in agreement with the lattice data at $T\lsim T_{\chi SR}$.

Let us finally turn to the third regime, namely above
$T_{\chi SR}$.
It has been shown by Prakash and Zahed \cite{prakash} that with increasing
temperature, the susceptibility goes to its perturbative value which can
be calculated with perturbative gluon-exchanges.
The argument is made with the dimensional reduction at asymptotic temperatures,
but it seems to apply even at a temperature slightly above $T_{\chi SR}$.
We will later
make a conjecture why this is reasonable. We shall schematize
the Prakash-Zahed argument using the dimensionally reduced model of Koch
{\etal}\ \cite{KSBJ} which hints at the onset of the Georgi vector limit.
In this model which exploits the ``funny space" obtained by interchanging
$z$ and $t$ (somewhat like ``moduli transform"), the helicity-zero state of
the $\rho$ meson is found to come out degenerate with the pion while the
helicity $\pm$ states are degenerate with each other. In finite temperature,
$z$ replaces $T$, so asymptotically in $T$, the configuration space with the
new $z$ becomes 2-dimensional with $x$ and $y$. The $\rho$ meson has gone
massless and behaves like a (charged) photon with helicities $\pm$ 1
perpendicular to the plane. The helicity-zero state originating from
the longitudinally polarized component of the $\rho$ before it went massless
now behaves as an isotriplet scalar. We identify this with the scalar $S(x)$
described above, a realization of the Georgi vector symmetry.

Let us assume then that the vector mesons have decoupled with $g=0$. Going to
the perturbative picture with quark-gluon exchanges, we take
one-gluon-exchange potential of Koch {\etal},
\be
V(r_t)=\frac{\pi}{m^2}\frac 43 \bar{g}^2 T
\sigma_{z,1}\sigma_{z,2}\delta (r_t)
\label{V}
\ee
with $\bar{g}$ the color gauge coupling and $\delta (r_t)$ is the
$\delta$-function in the two-dimensional reduced space. Here $m=\pi T$ is
the chiral mass of quark or antiquark as explained in \cite{KSBJ}.
Possible constant terms that can contribute to eq.(\ref{V}) will be ignored
as in \cite{KSBJ}.
In order to evaluate the expectation value of the $\delta (r_t)$, we note that
the helicity-zero
$\rho$-meson wave function in two dimensions is well approximated by
\be
\psi_\rho\approx N e^{-r_t/a}
\ee
with $a\approx \frac 23$ fm and the normalization
\be
N^2=\frac{2}{\pi a^2}.
\ee
For the helicity $\pm 1$ $\rho$-mesons, $\sigma_{z,1}\sigma_{z,2}=1$,
so we find that the expectation value of $V$ is
\be
\langle V\rangle=\frac 83 \frac{\bar{g}^2 T}{\pi^2 T^2 a^2}.
\ee

Now summing the ladder diagrams to all orders, we get
\be
\frac{\chi}{\chi_0}=\left(1+\frac{\langle V \rangle}{2\pi T}\right)^{-1},
\label{ratio}
\ee
where the energy denominator $2\pi T$ corresponds to the mass of a pair
of quarks.

The lattice calculations \cite{gottliebchi} use $6/\bar{g}^2=5.32$
which would give $\alpha_s=0.07$ at scale of $a^{-1}$ where $a$ is the lattice
spacing. (The relevant scale may be more like $2\pi/a$.) Calculations
use  4 time slices, so the renormalized $\bar{g}$ is that appropriate
to $a^{-1/4}$. Very roughly we take this into account by multiplying the
above $\alpha_s$ by $\ln 4^2$; therefore using $\alpha_s\cong 0.19$.
With this $\alpha_s$ and the above wave function, we find
\be
\frac{\chi (T_{\chi SR}^+)}{\chi_0 (T_{\chi SR}^+)}\approx 0.68.
\label{pertchi}
\ee
This is just about the ratio obtained above $T_{\chi SR}$
in the lattice calculations. Remarkably the perturbative result
(\ref{pertchi}) above $T_c$ matches smoothly onto
the HGS prediction (\ref{HGSchi}) just below $T_c$. Neglecting logarithmic
dependence of the gauge coupling constant, eq. (\ref{ratio}) can be
written as
\be
\frac{\chi}{\chi_0} (T)\approx \frac{1}{1+0.46 (T_c/T)^2}
\ee
which follows closely the lattice gauge results of
Gottlieb {\etal}\ \cite{gottliebchi}. We consider this an
indication for the Georgi vector symmetry in action, with the
induced flavor gauge symmetry in the hadronic sector ceding to the
fundamental color gauge symmetry of QCD in the quark-gluon sector.

We should remark that to the extent that the screening mass obtained in
\cite{KSBJ} $m_\pi=m_S\approx 2\pi T$ is consistent with two non-interacting
quarks and that the corresponding wave functions obtained therein are
the same for the pion $\pi$ and the scalar $S$, we naturally expect the
relation
(\ref{equal}) to hold. A short calculation
of the matrix element of the axial current $A_\mu$
with the $\pi$ wave function in the
``funny space" gives, for large $T$ \footnote{We denote
the constant by $\tilde{f}_\pi$
to distinguish it from the physical pion decay constant $f_\pi$.}
\be
\tilde{f}_\pi \sim c\sqrt{\bar{g}} T\label{decay}
\ee
where $c$ is a constant $<< 1$ and $\bar{g}$ is
the color gauge coupling constant.

\subsubsection*{``Cool" kaons in heavy-ion collisions}
\indent

The vanishing of the hidden gauge coupling $g$ can have a dramatic effect
on the kaons produced in relativistic heavy-ion collisions. In particular,
it is predicted that the kaons produced from quark-gluon plasma would have a
component that has
a temperature much lower than that for other hadrons. This
scenario may provide an explanation of the recent preliminary
data\cite{stachel} on the $14.6$ GeV collision (experiment E-814)
\be
^{28}{\rm Si} + {\rm Pb}\rightarrow K^+ (K^-) + X
\ee
which showed cool components with effective
temperature of 12 MeV for $K^+$ and 10 MeV for $K^-$, which cannot be
reproduced in the conventional scenarios employed in event generators.
The latter give kaons of effective temperature $\sim 150$ MeV.

There are two points to keep in mind in understanding what is happening
here. Firstly, the Brookhaven AGS experiments determined the freeze-out --
the effective decoupling in the sense of energy exchange of pions and
nucleons -- at $T_{fo}\approx 140$ MeV\cite{BSW}\footnote{The original
determination of $T\gsim 150$ MeV from the
ratio of isobars to nucleons by Brown, Stachel and Welke \cite{BSW}
was corrected about 10 MeV downward by taking effects such as the finite
width of the isobar into account.}. This is essentially the same as
the chiral transition temperature measured in lattice gauge calculations
\cite{lattice}. This suggests that the freeze-out for less strongly interacting
particles other than
the pion and the nucleon is at a temperature higher than $T_{\chi SR}$ and
that the pion and nucleon freeze out at about $T_{\chi SR}$. This means
that interactions in the interior of the fireball will be at temperature
greater than $T_{\chi SR}$. At this temperature, the vector coupling
$g$ would have gone to zero, so the Georgi vector limit would be
operative were it to be relevant.
The second point is that the fireball must expand slowly.
The slow expansion results because the pressure in the region
for some distance above $T_{\chi SR}$ is very low \cite{kochbrown}, the
energy in the system going into decondensing gluons rather than giving
pressure.
This results in an expansion velocity of $v/c\sim 0.1$.
In the case of 15 GeV/N Si on Pb transitions, the fireball has been measured
\cite{braun} through Hanbury-Brown-Twiss correlations of the pions to increase
from a transverse size of $R_T (Si)=2.5$ fm to $R_T=6.7$ fm, nearly a factor
of 3, before pions freeze out. With an expansion velocity of $v/c\sim 0.1$,
this means an expansion time of $\sim 25 - 30$ fm/c. (The full expansion
time cannot be measured from the pions which occur as a short flash at the
end.)

In a recent paper, V. Koch\cite{koch}
has shown that given a sizable effective attractive
interaction between the $K^+$ and the nucleon at the freeze-out phase,
a cool kaon component can be reproduced in the conditions specified above.
We argue now that such an attractive
interaction can result if the Georgi vector limit is realized.

The description by chiral perturbation theory\cite{knpw,LJMR,LBR,LBMR}
of kaon nuclear interactions and kaon condensation in
dense nuclear matter has shown that three mechanisms figure prominently
in kaon-nuclear processes at low energy: (1) the $\omega$ meson exchange
giving rise to repulsion for $K^+ N$ interactions and attraction
for $K^- N$; (2) the ``sigma-term" attraction for both $K^\pm N$:
(3) the repulsive ``virtual pair term." In effective chiral
Lagrangians, the first takes the form, $\sim \pm
\frac{1}{f^2} K^\dagger \del_\mu K \bar{N} \gamma^\mu N$ for $K^\pm$,
the second $\sim \frac{\Sigma_{KN}}{f^2} K^\dagger K \bar{N} N$
and the third term $\sim (\del_\mu K)^2 \bar{N}N$\footnote{The $\Lambda (1405)$
driven by the vector-exchange term plays an important role in $K^-p$ scattering
but an irrelevant role in kaon condensation and no role at all for
$K^+ N$ processes.}. Roughly the vector-exchange gives the repulsive
potential\footnote{Note that by $G$-parity, this potential turns attractive
for $K^- N$.}
\be
V_{K^+ N}\cong \frac 13 V_{NN}\cong 90\ {\rm MeV}\,\frac{\rho}{\rho_0}
\label{repulsion}
\ee
where $\rho_0$ is nuclear matter density. This term is proportional to
the hidden gauge coupling $g^2$. One can estimate the scalar attraction
by the ``sigma term"
\be
S_{K^+ N}\approx -\frac{\Sigma_{KN} \langle \bar{N}N\rangle}
{2 m_K f^2}\cong -45\ {\rm MeV}\,\frac{\rho_s}{\rho_0}
\label{attraction}
\ee
where $\rho_s$ is the scalar density and $\Sigma_{KN}$ is the $KN$ sigma
term.  Being $G$-parity invariant, this remains attractive for $K^- N$
interactions. The virtual pair term
(proportional to $\omega^2$ where $\omega$ is
the kaon frequency) -- which is
related to Pauli blocking -- removes, at zero temperature, about 60 \%
of the attraction (\ref{attraction}). At low temperature, the net effect is
therefore highly repulsive for $K^+ N$ interactions.

It is easy to see what happens as $T\rightarrow T_{\chi SR}$. First of all,
part of the virtual pair repulsion gets ``boiled" off as discussed in
\cite{BKR}. What is more
important, if the Georgi vector limit is relevant, then
the vector mesons decouple with $g\rightarrow 0$, killing off the repulsion
(\ref{repulsion}). As a consequence, the residual attraction from the
scalar exchange remains. The calculation of Koch\cite{koch} supports this
scenario. Given that the vector coupling is absent, one can see that
both $K^+$ and $K^-$ will have a similar cool component.
\subsubsection*{Instanton-molecule model for chiral restoration}
\indent

As a final case, we mention a microscopic model that seems to
realize the Georgi vector symmetry at high temperature.

In a model where the chiral phase transition is described as a change
in the instanton liquid from a randomly distributed phase at low temperature
to an instanton-anti-instanton molecular phase above $T_{\chi SR}$, it has been
observed\cite{schafer} that the molecules get polarized in the time direction
and the interactions in the pion and the longitudinal vector channel
become identical. This leads to the degeneracy of the triplets of
$\pi$ and $\rho_\parallel$ which may be identified with the scalar $S$.
The interaction in the longitudinal vector-meson
channel becomes equally strong as attraction in
the scalar-pseudoscalar channel, while
transversely polarized vector mesons have no interaction.
If one assumes that the polarized molecules are the dominant agent for
interactions above $T_{\chi SR}$, then one can see that all coupling
constants in an NJL-type effective Lagrangian so generated could be
specified in terms of a {\it single} coupling constant, implying the swelling
of the symmetry in a manner closely paralleling the Georgi vector symmetry.
In this case, the restored
symmetry is $U(2)\times U(2)$ since the axial $U(1)_A$
is also supposed to be restored. Perturbative QCD effects are not expected
to modify this symmetry structure but it is not clear that no other
non-perturbative effects can enter to upset this. Nonetheless this is
a microscopic picture consistent with the Georgi vector symmetry.

\subsubsection*{Discussions}
\indent

We have seen that the behavior of the quark number
susceptibility with temperature, as calculated in lattice gauge calculations,
shows that the hadronic vector coupling disappears as $T$ moves upwards
through $T_{\chi SR}$, and that the perturbative color gluon exchange
describes the susceptibility well above $T_{\chi SR}$, as argued by
Prakash and Zahed\cite{prakash}. Somewhat surprising is the fact that
the perturbative description, which gives a $1/T^2$ behavior in the difference
$\chi (T)-\chi (0)$ between the susceptibility and that for free quarks,
sets in just above $T_{\chi SR}$; {\ie}, for this purpose, asymptotia is
$T \gsim T_{\chi SR}$. Note that the screening mass of the
$\rho$-meson goes to $2\pi T$, its asymptotic value, as soon as $T$ reaches
$T_{\chi SR}$. Why is asymptotia reached so soon?

We suggest that the relevant parameter for reaching asymptotia for this
particular quantity is $T/m_{dyn}$; {\ie}, the ratio of temperature to
dynamically generated mass. The temperature $T_{\chi SR}$ need not be
very large as long as $m_{dyn}\rightarrow 0$ at $T_{\chi SR}$ in order to reach
asymptotia with this parameter.

While other viable explanations may be found in the future, as far as we know,
the soft kaons found in the E-814 experiment can be explained only if the
vector-meson coupling is essentially absent at the relevant temperature, and
this is a strong support for the Georgi picture in which the hidden gauge
coupling constant ``melts" at the temperature $T_{\chi SR}$.
Now the gauge symmetry in the hidden local symmetry
scheme is an ``induced" gauge symmetry lodged in hadronic variables.
In this sector, the fundamental color gauge symmetry is not
visible. It is the induced flavor one that is seen. What we observe is
then that as $T$ goes towards $T_{\chi SR}$, the induced gauge symmetry
gives way to the fundamental gauge symmetry. What is surprising is that
this changeover seems to take place suddenly, with increasing temperature,
the effective hadronic gauge symmetry applying for $T < T_{\chi SR}$,
and the fundamental color gauge symmetry being realized
perturbatively for $T > T_{\chi SR}$.

Finally we would like to suggest that while the Georgi vector limit
is relevant to the chiral symmetry restoration at high temperature and/or
high density, the
``mended symmetry" with $\kappa=0$ and $g\neq 0$  (with $m_\rho\neq 0$)
may also be relevant
for nuclear physics at temperatures and matter densities below the critical
values. As pointed out by Beane and van Kolck\cite{beane}, it is the
existence of a symmetry of this type (involving a light dilaton)
that allows one to linearize the non-linear chiral
Lagrangian of current algebra to the linear sigma model at some shorter
length scale, {\eg}, in nuclear medium. As discussed elsewhere\cite{newbr}
it is this mechanism that allows a separation into two components of the
scalar $\chi$ field that enters into the trace anomaly of QCD and gives
rise to the medium-scaling (so-called ``BR scaling") proposed in
ref.\cite{br91} of effective chiral Lagrangians in nuclear medium.
For more discussion on this issue, see refs.\cite{newbr,mrelaf}.

\subsection*{Acknowledgments}
\indent

We are grateful for discussions with Tetsuo Hatsuda, Volker Koch, Maciej Nowak,
Koichi Yamawaki and Ismail Zahed.


\begin{thebibliography} {000}
\bibitem{georgi} H. Georgi, \np \ {\bf B331} (1990) 311.
\bibitem{weinbergMS} S. Weinberg, \prl \ {\bf 65} (1990) 1177;
``Unbreaking symmetries," in Festschrift for Abdus Salam, to be published.
\bibitem{bando} M. Bando, T. Kugo and K. Yamawaki, \PR \ {\bf 164} (1988) 217.
\bibitem{harada} M. Harada and K. Yamawaki, \pl \ {\bf B297} (1992) 151.
\bibitem{haradaPRL} M. Harada, T. Kugo and K. Yamawaki, \prl \ {\bf 71} (1993)
1299.
\bibitem{gottliebchi} S. Gottlieb, W. Liu, D. Toussaint, R.C. Renken
and R.L. Sugar, \prl \ {\bf 59} (1987) 2247.
\bibitem{kunihiro} T. Kunihiro, \pl \ {\bf B271} (1991) 395.
\bibitem{prakash} M. Prakash and I. Zahed, \prl \ {\bf 69} (1992) 3282.
\bibitem{KSBJ} V. Koch, E.V. Shuryak, G.E. Brown and A.D. Jackson,
\pr \ {\bf D46} (1992) 3169; \pr \ {\bf D47} (1993) 2157 (E).
\bibitem{stachel} J. Stachel, \np \ {\bf A566} (1994) 183c.
\bibitem{BSW} G.E. Brown, J. Stachel and G.M. Welke, \pl \ {\bf B253} (1991)
19.
\bibitem{lattice} C. Bernard, M.C. Ogilvie, T.A. DeGrand, C. DeTar,
S. Gottlieb, A. Krasnitz, R.L. Sugar and D. Toussaint, \pr \ {\bf D45}
(1992) 3854.
\bibitem{kochbrown} V. Koch and G.E. Brown, \np \ {\bf A560} (1993) 345.
\bibitem{braun} P. Braun-Munzinger, Proc. NATO Adv. Study Ins., Bodrum,
Turkey, Oct. 1993, to appear.
\bibitem{koch} V. Koch, Stony Brook preprint SUNY-NTG 94-26; nucl-th/9405022.
\bibitem{knpw} D.B. Kaplan and A.E. Nelson, \pl \ {\bf B175} (1986) 57;
H.D. Politzer and M.B. Wise, \pl \ {\bf B273} (1991) 156.
\bibitem{LJMR} C.-H. Lee, H. Jung, D.-P. Min and M. Rho, \pl \ {\bf B326}
(1994) 14.
\bibitem{LBR} C.-H. Lee, G.E. Brown and M. Rho, ``Kaon condensation in
`nuclear star' matter," Phys. Lett., to be published.
\bibitem{LBMR} C.-H. Lee, G.E. Brown, D.-P. Min and M. Rho,
``An effective chiral Lagrangian approach to kaon-nuclear interactions:
Kaonic atom and kaon condensation," hep-ph/9406311, to be published.
\bibitem{BKR} G.E. Brown, V. Koch and M. Rho, \np \ {\bf A535} (1991) 701.
\bibitem{schafer} T. Sch\"{a}fer, E.V. Shuryak and J.J.M. Verbaarschot,
preprint  SUNY-NTG-94-24.
\bibitem{beane} S. Beane and U. van Kolck, \pl \ {\bf B328} (1994) 137.
\bibitem{newbr} G.E. Brown and M. Rho, ``Chiral restoration in hot and/or
dense matter," to appear.
\bibitem{br91} G.E. Brown and M. Rho, \prl \ {\bf 66} (1991) 2720.
\bibitem{mrelaf} M. Rho, Phys. Repts. {\bf 240} (1994) 1.

\end{thebibliography}
\end{document}